\documentclass[journal,twocolumn]{IEEEtran-v18}
\pdfoutput=1

\newlength{\figwidth}
\usepackage{color}
\definecolor{links}{rgb}{0.7,0,0}   % red
\definecolor{urls}{rgb}{0,0,0.8}    % blue
\definecolor{cites}{rgb}{0,0,0.8}   % blue

\usepackage[colorlinks,hyperindex,linkcolor=links,citecolor=cites,urlcolor=urls]{hyperref} % generates colored links in pdf file
 
\usepackage[nosort]{cite}        
\usepackage{url} 
\usepackage[intlimits]{amsmath}
\usepackage{bbm}
\usepackage{graphicx}
\usepackage{paralist}
\usepackage{fancyref}
\usepackage[stretch=16,shrink=16,step=4]{microtype}
\usepackage{vmr-symbols-vecbold}
\usepackage{standard-macros}
\usepackage{cases}

\usepackage{stfloats}

\usepackage{balance}

\makeatletter
\def\@IEEEinterspaceratioM{0.265}
\def\@IEEEinterspaceMINratioM{0.1651}
\def\@IEEEinterspaceMAXratioM{0.38}

\def\@IEEEinterspaceratioB{0.31}
\def\@IEEEinterspaceMINratioB{0.19}
\def\@IEEEinterspaceMAXratioB{0.38}
\@IEEEtunefonts
\makeatother
\begin{document}

\IEEEoverridecommandlockouts
% DRAFT
% to include revision information into the resulting PDF
%%%%\svnInfo $Id: dbs_it07.tex 2697 2009-08-04 10:43:20Z gdurisi $ 
%% add a PDFinfo field to store metadata in the output PDF

\title{Oversampling Increases the Pre-Log of\\Noncoherent Rayleigh Fading Channels}
\author{Meik~D\"orpinghaus,~\IEEEmembership{Member,~IEEE}, G\"unther~Koliander,~\IEEEmembership{Student~Member,~IEEE}, Giuseppe~Durisi,~\IEEEmembership{Senior~Member,~IEEE}, Erwin~Riegler,~\IEEEmembership{Member,~IEEE}, and Heinrich~Meyr,~\IEEEmembership{Life~Fellow,~IEEE}
\thanks{The work of M.\ D\"orpinghaus was supported in part by the Deutsche Forschungsgemeinschaft (DFG) under grant DO1568/1-1 and by the DFG in the framework of the Collaborative Research Center 912 ``Highly Adaptive Energy-Efficient Computing''. The work of G.\ Koliander and E.\ Riegler was supported by the WWTF under grant ICT10-066 (NOWIRE). The work of G.\ Durisi was supported in part by the Swedish Research Council under grant 2012-4571.}     
\thanks{M.~D\"orpinghaus is with the Vodafone Chair Mobile Communications Systems, Technische Universit\"at Dresden, 01062 Dresden, Germany (e-mail: meik.doerpinghaus@tu-dresden.de).}
\thanks{G.~Koliander is with the Institute of Telecommunications, Vienna University of Technology, 1040 Vienna, Austria (e-mail: guenther.koliander@nt.tuwien.ac.at).}
\thanks{G.~Durisi is with the Department of Signals and Systems, Chalmers University of Technology, 41296 Gothenburg, Sweden (e-mail: durisi@chalmers.se).}
\thanks{E.~Riegler is with the Department of Information Technology and Electrical Engineering, ETH Zurich, CH-8092 Zurich, Switzerland (e-mail: eriegler@nari.ee.ethz.ch).}
\thanks{H.~Meyr is an emeritus of the Institute for Integrated Signal Processing Systems, RWTH Aachen University, 52056 Aachen, Germany and is now a grand professor of the Center for Advancing Electronics Dresden (cfaed) at Technische Universit\"at Dresden, 01062 Dresden, Germany (e-mail: \mbox{meyr@iss.rwth-aachen.de).}}
%\thanks{Copyright (c) 2014 IEEE. Personal use of this material is permitted.  However, permission to use this material for any other purposes must be obtained from the IEEE by sending a request to pubs-permissions@ieee.org.}
}

\maketitle

%%%%%%%%%%%%%%%%
\begin{abstract}
We analyze the capacity of a continuous-time, time-selective, Rayleigh block-fading channel in the high signal-to-noise ratio (SNR) regime. 
The fading process is assumed stationary within each block and to change independently from block to block; furthermore, its realizations are not known {a priori} to the transmitter and the receiver (noncoherent setting).
A common approach to analyzing the capacity of this channel is to assume that the receiver performs matched filtering followed by sampling at symbol rate ({symbol matched filtering}). 
This yields a discrete-time channel in which each transmitted symbol corresponds to one output sample. 
Liang \& Veeravalli (2004) showed that the capacity of this discrete-time channel grows logarithmically with the SNR, with a capacity {pre-log}  equal to $1-{Q}/{N}$. 
Here, $N$ is the number of symbols transmitted within one fading block, and  $Q$ is the rank of the covariance matrix of the discrete-time channel gains within each fading block. 
In this paper, we show that symbol matched filtering is {not} a capacity-achieving strategy for the underlying continuous-time channel.
Specifically, we analyze the capacity pre-log of the discrete-time channel obtained by {oversampling} the continuous-time channel output, i.e., by sampling it  faster than at symbol rate. 
We prove that by oversampling by a factor two one gets a capacity pre-log that is at least as large as $1-1/N$.
Since the capacity pre-log corresponding to symbol-rate sampling is $1-Q/N$, our result implies indeed that symbol matched filtering is not capacity achieving at high SNR.
\end{abstract}

%%%%%%%%%%%%%%%%%%%%%%%%%%%%%%%%%%
\section{Introduction} % (fold)
\label{sec:introduction}

We analyze the high signal-to-noise ratio (SNR) capacity of  a continuous-time, time-selective, Rayleigh block-fading  channel, whose realizations are unknown to the transmitter and the receiver (noncoherent setting). 
Computing the channel capacity in this scenario is relevant because the propagation environment is typically not known \emph{a priori} at the transceivers and needs to be estimated, for example through the transmission of pilot symbols.

Information-theoretic analyses of fading channels in the noncoherent setting are typically conducted starting from discrete-time models.
Marzetta and Hochwald~\cite{marzetta99-01a,hochwald00-03a} investigated the capacity of discrete-time Rayleigh block-fading channels where the channel remains constant for $N$ channel uses before changing to an independent realization.
For the case of single-input single-output (SISO) channels, they provided a closed-form characterization of capacity in the  high SNR regime in terms of both capacity \emph{pre-log}, i.e., the asymptotic ratio between capacity and the logarithm of SNR as SNR grows large, and \emph{second-order} term in the high-SNR capacity expansion. 
Zheng and Tse~\cite{zheng02-02a} (and, more recently, Yang \emph{et al.}~\cite{yang13-02a}) extended these results to the multiple-input multiple-output (MIMO) case, and provided a geometric interpretation of the problem of communicating over a Rayleigh block-fading channel whose realizations are unknown to the receiver. 

An alternative way to model channel variations in time is to assume that the fading gains evolve according to a discrete-time stationary process.
Focusing on this scenario, Lapidoth and Moser~\cite{lapidoth03-10a} proved that---for general fading distributions---capacity grows double-logarithmically with SNR whenever the fading process is \emph{regular}, i.e., when the present fading state cannot be inferred from the knowledge of arbitrarily many past fading states.
To get a more detailed understanding of the high SNR capacity for the case of regular fading, one has to study the second-order term in the high-SNR capacity expansion, the so called \emph{fading number}.
Its value has been recently characterized for several stationary discrete-time channel models~\cite{lapidoth03-10a,lapidoth06-02a,moser09-06a,koch09-05a}.

For the case of \emph{nonregular} Rayleigh fading, where nonregular means that the present fading state can be inferred from the knowledge of the past fading states, the high-SNR capacity behavior depends on the support of the power spectral density (PSD) of the fading process. 
Specifically, the capacity pre-log is given by the Lebesgue measure of the set of frequencies at which the PSD vanishes~\cite{lapidoth05-07a}.
Moreover, it is shown in~\cite{koch09-05a} that Rayleigh fading yields the smallest capacity pre-log among all stationary and ergodic fading processes with a given PSD whose law has no mass point at zero.

Departing from the approaches followed in~\cite{marzetta99-01a,hochwald00-03a,zheng02-02a,lapidoth03-10a}, Durisi \emph{et al.}~\cite{durisi12-10a} analyzed the high-SNR capacity of \emph{continuous-time} Rayleigh-fading time-frequency selective channels.
However, the generality of the model considered by the authors prevented them to obtain  a closed-form expression for the high-SNR capacity.
In \cite{liang04-09a}, the high-SNR capacity of a continuous-time, time-selective, frequency-flat, Rayleigh block-fading channel is studied.
This channel model, which is sometimes referred to as \emph{correlated block-fading} model \cite{morgenshtern13-07a,koliander13-10a},  is a generalization of the standard block-fading model as it allows the fading process to change in a stationary manner within each block. 
To investigate its high-SNR capacity, the authors in~\cite{liang04-09a} approximate the underlying continuous-time channel by a discrete-time channel obtained by performing matched filtering at the receiver followed by sampling at symbol rate (\emph{symbol matched filtering}).
For the SISO case, it is shown that the capacity pre-log is given by $1-Q/N$ where $N$ is the number of symbols transmitted within one fading block of duration $T$ seconds, and $Q$ is the rank of the covariance matrix of the channel fading gains within one fading block.
The rank $Q$ is related to the maximum Doppler frequency $\nu_{\textrm{max}}$, namely, $Q=2\floor{\nu_{\textrm{max}}T}+1$.
Equivalently, the capacity pre-log is given by the number of zero eigenvalues of the $N\times N$ covariance matrix of the discrete-time channel fading process within one fading block, normalized with respect to $N$.
This result has the same flavor as the one obtained in~\cite{lapidoth05-07a} for the stationary case (see~\cite{jindal10-10a} for a comparison between stationary and block-fading models). 
%

%It has to be mentioned that except of \cite{lapidoth03-10a,lapidoth06-02a,moser09-06a} all of the referenced papers consider Rayleigh fading channels exclusively. 

To summarize, high-SNR capacity characterizations in terms of pre-log for the noncoherent setting are available only  for discrete-time channels~\cite{marzetta99-01a,hochwald00-03a,zheng02-02a,lapidoth03-10a,lapidoth05-07a} or for the case when continuous-time, time-selective, frequency-flat channels are discretized using the symbol matched filtering approach~\cite{liang04-09a}.
However, as we shall detail below, the multiplication of a continuous-time input signal by the time-selective, frequency-flat fading process yields in general a bandwidth expansion.
So symbol matched filtering is not necessarily optimal, because the resulting discretized channel output is not a sufficient statistics.

\subsubsection*{Contributions} % (fold)
\label{sec:contributions_}
Focusing on the time-selective, frequency-flat, Rayleigh block-fading model introduced in~\cite{liang04-09a}, we investigate whether symbol matched filtering is optimal from a pre-log point of view.
Intuitively, to achieve capacity it appears necessary to sample the continuous-time received signal so as to obtain a sufficient statistics for the detection of the transmitted signal~\cite[Ch.~4.2.4]{Meyr1998},~\cite[Ch.~8]{gallager68a},~\cite{gelfand57-a}.
Assume for example that the channel input signal has a (one-sided) bandwidth of $W=1/(2T_{S})$ where $T_{S}$ denotes the duration of one symbol. 
If the channel fading process has a bandwidth (maximum Doppler spread) of $\nu_{\textrm{max}}$, the bandwidth of the noiseless received signal  is  $W+\nu_{\textrm{max}}$.
Thus, in order to fulfill the Nyquist condition and to be able to reconstruct the received signal from its samples, 
one has to sample at a rate not smaller than $2(W+\nu_{\textrm{max}})$.
Under the assumption that $\nu_{\textrm{max}}< W$, which holds for most wireless communication systems, sampling the received signal for example at twice the symbol rate, i.e., oversampling by a factor $2$, yields a sufficient statistics. 

The question is whether oversampling actually yields a higher mutual information at high SNR. % even when $\nu_{\textrm{max}}$ is much smaller than $W$ (this often occur in practice).
In this paper, we show that this is indeed the case.
Specifically, for the continuous-time channel model considered in~\cite{liang04-09a}, we prove that the discrete-time channel obtained by  oversampling the output signal by a factor $2$ has a capacity pre-log that is at least as large as $1-{1}/{N}$ for every $Q<N$. 
This capacity pre-log lower bound is independent of the rank $Q$ of the channel covariance matrix of the individual fading blocks.
In contrast---as already discussed---the discrete-time channel obtained with symbol rate sampling has a capacity pre-log equal to $1-{Q}/{N}$~\cite[Th.~1]{liang04-09a}.
Hence, we conclude that the capacity pre-log of the underlying time-selective, frequency-flat, Rayleigh block-fading channel is at least $1-1/N$, and that
symbol matched filtering is not capacity achieving at high SNR. 
%However, since the proof of the lower bound for the capacity pre-log in the present paper is based on rectangular time-limited transmit pulses, we do not claim that sampling with twice the symbol rate, i.e., oversampling by a factor of $2$, provides a sufficient statistics. In a forthcoming paper, we address the case of bandlimited transmit pulses which are sampled at a rate which provides a sufficient statistics.

\subsubsection*{Proof Techniques} % (fold)
\label{par:proof_techniques}
For technical reasons related to the operational definition of capacity for continuous-time channels~\cite[Ch.~8]{gallager68a}, \cite{wyner66-03a}, we do not consider the transmission of bandlimited signals; rather, we choose signals that are a linear combination  of time-shifted, time-limited transmit pulses.\footnote{As pointed out in~\cite[p.~364]{wyner66-03a}, the notion of rate, which is central to the definition of capacity, has only a limited operational meaning when the transmitted waveforms are strictly bandlimited, and, hence, of infinite time duration.}
Under this assumption,  sampling the received signal at twice the symbol rate  does not yield a sufficient statistic because the transmitted signal is not bandlimited.
As we shall see, oversampling  yields nevertheless a capacity pre-log increase.

The techniques used to establish the lower bound on the capacity pre-log for the oversampled case proposed in this paper are similar to the ones 
used in~\cite{morgenshtern13-07a,koliander13-10a}. 
In~\cite{morgenshtern13-07a}, the authors consider the discrete-time correlated Rayleigh block-fading model that results from performing symbol matched filtering on the continuous-time channel model proposed in~\cite{liang04-09a} and show that by adding sufficiently many receive antennas one can lift the capacity pre-log from $1-Q/N$ to $1-1/N$. 
This shows that the pre-log penalty $Q/N$ due to lack of \emph{a priori} channel knowledge (recall that for the case of perfect channel knowledge at the receiver the capacity pre-log is $1$) can be made small by adding additional receive antennas, i.e., by ``oversampling in space''.
One fundamental difference between our setup and the one considered in~\cite{morgenshtern13-07a,koliander13-10a} is that adding more antennas increases the number of fading parameters to estimate. 
As we shall see, this does not occur in the oversampled discrete-time model considered in the present paper.

% paragraph proof_techniques (end)

\subsubsection*{Related Results} % (fold)
\label{par:related_results}

% paragraph related_results (end)
The fact that oversampling increases the rates achievable at high-SNR has  been recently observed in the context of phase-noise channels \cite{ghozlan13-07a}. Specifically,  it is shown in~\cite{ghozlan13-07a} for the case of Wiener phase noise that, by oversampling the output signals by a factor that grows with the square-root of SNR, one can achieve rates that grow logarithmically with the SNR and a pre-log no smaller than~$1/2$. 
In contrast, symbol matched filtering yields only a double-logarithmic growth of capacity with the SNR. 
Oversampling increases the capacity also for the case of AWGN channels with output quantization, both at low SNR~\cite{koch10-11a} and in the limiting case of no additive noise~\cite{Gilbert93,Shamai94}.

\subsubsection*{Notation} 
Calligraphic letters like $\mathcal{M}$ denote sets, and $|\mathcal{M}|$ stands for the cardinality of the set $\mathcal{M}$. 
Boldface lower case letters such as $\mathbf{a}$ and $\mathbf{b}$ and upper case letters such as $\mathbf{A}$ and $\mathbf{B}$ denote vectors and matrices, respectively. 
Random quantities are denoted by sans serif letters, e.g., $\bm{\mathsf{A}}$ is a random matrix and $\bm{\mathsf{a}}$ is a random vector. 
The notation $[M:N]$ is used to indicate the set $\{n\in\mathbb{N}: M\le n\le N\}$ for $M,N\in \mathbb{N}$.  
%The element on the $k$th row and $l$th column of the matrix $\mathbf{A}$ is denoted by $[\mathbf{A}]_{k}^{l}$.  
%Furthermore, $[\mathbf{A}]_{\mathcal{I}}^{\mathcal{J}}$ is a matrix containing only the rows with indices in $\mathcal{I}\subseteq [1:M]$ and the columns with indices in $\mathcal{J}\subseteq [1:N]$ of the $M\times N$ matrix $\mathbf{A}$. 
We write $[\mathbf{A}]_{\mathcal{I}}$ to denote a submatrix of $\mathbf{A}$ containing only the rows with indices in $\mathcal{I}\subseteq [1:M]$ %and, likewise, $[\mathbf{A}]^{\mathcal{J}}$ is a submatrix of $\mathbf{A}$ containing only the columns with indices in $\mathcal{J}$. 
of the $M\times N$ matrix $\mathbf{A}$. For the vector $\mathbf{a}$, $[\mathbf{a}]_{\mathcal{I}}$ is a subvector containing the elements with indices in $\mathcal{I}$. 
In addition, $\mathbf{I}_{N}$ is the identity matrix of dimension $N\times N$, and $\mathbf{0}_{N\times M}$ is the all-zero matrix of size  $N\times M$. 
The operator $\diag(\mathbf{a})$ generates a square diagonal matrix with the elements of $\mathbf{a}$ on its main diagonal; 
$\det(\mathbf{A})$ stands for the determinant of $\mathbf{A}$ and we abbreviate the absolute value of the determinant of $\mathbf{A}$ with $|\mathbf{A}|$. 
The superscript $^{T}$,~$^{H}$, and~$^{*}$ denote transpose, Hermitian transpose, and complex conjugate, respectively. 
For $x\in\mathbb{R}$, we define $\floor{x}=\max\{m\in\mathbb{Z}:m\le x\}$. 
The notation $f(\cdot)=\mathcal{O}(g(\cdot))$ for two functions $f(\cdot)$ and $g(\cdot)$ means that $\lim_{u\rightarrow\infty}|f(u)/g(u)|$ is upper-bounded by a constant. 
The expectation operator is denoted by $\opE[\cdot]$. 
The $\rect(\cdot)$ function is defined as 
\begin{IEEEeqnarray}{rCL}
\rect(t)&=&\left\{\begin{array}{ll}
1, & \textrm{if } |t|\le {1}/{2}\\
0, & \textrm{otherwise}
\end{array}\right. 
\end{IEEEeqnarray}
and the $\sinc(\cdot)$ function is defined as
\begin{IEEEeqnarray}{rCL}
\sinc(x)&=&\left\{\begin{array}{ll}
\sin(\pi x)/(\pi x), & \textrm{if } x\ne 0\\
1, & \textrm{if } x=0.
\end{array}\right. 
\end{IEEEeqnarray}
%
%The convolution of two functions $f(t)$ and $g(t)$ is written as $(g\star f)(t)$. 
Finally, $\mathcal{CN}(\mathbf{0},\mathbf{C})$ denotes the probability distribution of a proper complex jointly Gaussian random vector with zero mean and %mean $\mathbf{\mu}$ and 
covariance matrix $\mathbf{C}$.

\section{System Model} % (fold)
\label{sec:system}
We consider the continuous-time, time-selective, Rayleigh-fading channel
\begin{IEEEeqnarray}{rCL}\label{eq:continuous-time-IO}
  %\IEEEeqnarraymulticol{3}{l}{...}
  % a & = & b +c
  \mathsf{y}(t)=\mathsf{h}(t)\mathsf{x}(t)+\mathsf{w}(t).
\end{IEEEeqnarray}
Here, $\mathsf{h}(t)$ is the channel fading process, $\mathsf{x}(t)$ is the transmit signal, $\mathsf{w}(t)$ is the additive white Gaussian noise process, and $\mathsf{y}(t)$ denotes the channel output. All these random quantities are complex.

We restrict ourselves to transmit signals of the form
  \begin{IEEEeqnarray}{rCL}
    \mathsf{x}(t)&=&\sum_{l=1}^{\infty}\sqrt{\rho}\ \mathsf{x}_l\ p(t-(l-1)T_{S})\label{eq:TxSignal}
  \end{IEEEeqnarray}
where in (\ref{eq:TxSignal}) the pulse $p(t)$ has unit energy and $p(t)=0$ if $t\notin (0,T_{S})$, with $T_{S}$ being the symbol duration, i.e., the measure of the support of $p(t)$. 
For simplicity, in the following we assume that $p(t)$ is a rectangular pulse, i.e.,\footnote{Recall that our aim is to establish an achievability result, i.e., a lower bound on the capacity pre-log. Hence, we are allowed to select a specific pulse shape. The choice of a rectangular pulse is convenient because it yields simpler mathematical expressions.
In practice, pulses with lower side lobes in frequency are preferable. Although our proof can be generalized to a larger family of pulse shapes, we decided to omit this extension because it is rather technical and may obfuscate the actual contribution of the paper.} 
\begin{IEEEeqnarray}{rCL}
  %\IEEEeqnarraymulticol{3}{l}{...}
  % a & = & b +c
  p(t)=\frac{1}{\sqrt{T_{S}}}\rect\lefto(\frac{t-{T_{S}}/{2}}{T_{S}}\right).
\end{IEEEeqnarray}

We also assume that the additive noise process $\mathsf{w}(t)$ is white zero-mean proper complex Gaussian.
The channel fading process $\mathsf{h}(t)$ is assumed zero-mean proper complex Gaussian, and stationary.
Hence, its dynamics are fully described by the correlation function 
\begin{IEEEeqnarray}{rCL}
r_{\mathsf{h}}(\tau)&=&\opE\left[\mathsf{h}(t+\tau)\mathsf{h}^{*}(t)\right]\label{CorrFunctio}
\end{IEEEeqnarray}
or, equivalently, by the PSD 
\begin{IEEEeqnarray}{rCL}\label{eq:PSD}
S_{\mathsf{h}}(\nu)&=&\int_{-\infty}^{\infty}r_{\mathsf{h}}(\tau)e^{-j2\pi \tau\nu}\mathrm{d}\tau.
\end{IEEEeqnarray}

Because the velocity of the transmitter, the receiver, and of the objects in the propagation environment are limited, it is reasonable to assume that $S_{\mathsf{h}}(\nu)$ has bounded support, say $[-\nu\sub{max},\nu\sub{max}]$, where $\nu\sub{max}$ is the maximum Doppler shift. 
This means that $\mathsf{h}(t)$ is a bandlimited process.

Large-scale effects involving changes not only in the phase, but also in the amplitude and the delay associated to each propagation path, may yield abrupt changes in the fading process. 
Following~\cite{liang04-09a}, we model these changes by assuming that stationarity, i.e.,~\eqref{CorrFunctio}, holds only over a time interval of length $T$ seconds. 
For mathematical tractability, but without losing the main features of the underlying physical process, we assume that the fading process takes on independent and identically distributed realizations across blocks of $T$ seconds.
Note that the interval length  $T$ is typically much larger than the coherence time $T\sub{coh}=1/(2\nu\sub{max})$ of the channel, which characterizes the time interval over which $\mathsf{h}(t)$ does not change significantly.
To summarize, we model $\mathsf{h}(t)$ as a block-memoryless process that satisfies~\eqref{CorrFunctio} within each block of $T$ seconds.
Over one such block, say the block $[0,T]$, we can express $\mathsf{h}(t)$ using the following series expansion
\begin{IEEEeqnarray}{rCL}
{\mathsf{h}}(t)&=&\sum_{m=-\infty}^{\infty}\mathsf{s}_{m}e^{j2\pi m\frac{t}{T}}, \quad t\in[0,T] \label{FourSerExp}
\end{IEEEeqnarray}
with
\begin{IEEEeqnarray}{rCL}
\mathsf{s}_{m}&=&\frac{1}{T}\int_{0}^{T}\mathsf{h}(t)e^{-j2\pi m\frac{t}{T}}\mathrm{d}t\label{FourCoeff}
\end{IEEEeqnarray}
and where the equality in~\eqref{FourSerExp} holds in mean-square sense.
Because $\mathsf{h}(t)$ is zero-mean proper complex Gaussian, the coefficients $\{\mathsf{s}_{m}\}$ are also zero-mean proper complex Gaussian. Furthermore, their cross-correlation is given by
\begin{IEEEeqnarray}{rCL}
\opE[\mathsf{s}_{m}\mathsf{s}_{n}^{*}]&=&\frac{1}{T}\int_{0}^{T}\left(\frac{1}{T}\int_{-\alpha}^{T-\alpha}r_{\mathsf{h}}(\tau)e^{-j2\pi m\frac{\tau}{T}}\mathrm{d}\tau\right. \nonumber\\&&\qquad\quad \left. \vphantom{\int_{-\alpha}^{T-\alpha}}\times e^{-j 2\pi m\frac{\alpha}{T}}\right)e^{j2\pi n\frac{\alpha}{T}}\mathrm{d}\alpha .\label{Approx_m2}
\end{IEEEeqnarray}
Under the assumption that the coherence time $T\sub{coh}$ of the fading process is much smaller than $T$ we can approximate $\opE[\mathsf{s}_{m}\mathsf{s}_{n}^{*}]$ as follows
\begin{IEEEeqnarray}{rCL}
\opE[\mathsf{s}_{m}\mathsf{s}_{n}^{*}]&\approx&
\left\{\begin{array}{ll}
\frac{1}{T}S_{\mathsf{h}}\lefto(\frac{m}{T}\right)& \textrm{for } m=n\\
0&\textrm{otherwise}
\end{array}\right.\label{SamlesofPSD}
\end{IEEEeqnarray}
which implies that the coefficients $\mathsf{s}_{m}$ can be assumed to be independent provided that $T\gg T\sub{coh}$.\footnote{The approximation~\eqref{SamlesofPSD} holds with equality for $T$-periodic correlation functions $r_{\mathsf{h}}(\tau)$.}

Since $\mathsf{h}(t)$ is bandlimited with (one-sided) bandwidth $\nu\sub{max}$, the series expansion in~\eqref{FourSerExp} can be well-approximated by keeping only a finite number of terms, i.e.,
\begin{IEEEeqnarray}{rCL}
{\mathsf{h}}(t)
&\approx&\sum_{m=-M}^{M}\mathsf{s}_{m}e^{j2\pi m\frac{t}{T}}, \quad t\in[0,T]\label{ChannelApproxExpans}
\end{IEEEeqnarray}
with $M=\floor{T\nu\sub{max}}$. The expansion on the right-hand side (RHS) of~\eqref{ChannelApproxExpans} implies that the fading process $\mathsf{h}(t)$ within the interval $[0,T]$ can be well-approximated by $Q=2M+1$ \emph{independent} proper complex Gaussian random variables, where  independence results from the approximation~\eqref{SamlesofPSD}. 
Since we assume that the fading process is memoryless across blocks of $T$ seconds, the entire fading process is characterized by specifying $Q$ independent parameters \emph{per block}.
It will turn out convenient to introduce  the normalized Fourier coefficients $\left\{\hat{\mathsf{s}}_m=\mathsf{s}_m/\sqrt{\frac{1}{T}S_{\mathsf{h}}\lefto(\frac{m}{T}\right)}\right\}_{m=-M}^{M}$, which are by construction \iid $\jpg(0,1)$-distributed.
In the remainder of the paper, we shall model the fading channel ${\mathsf{h}}(t)$ over each block using~\eqref{SamlesofPSD} and~\eqref{ChannelApproxExpans}.

\section{Symbol Matched Filtering}\label{sec:symbol matched filtering approach}
The symbol matched filtering approach, which involves 
filtering the received signal $\mathsf{y}(t)$ in~\eqref{eq:continuous-time-IO} with $p^{*}(-t)$ and then sampling at symbol rate $1/T_{S}$ yields for each fading block the following discrete-time input-output relation:
\begin{IEEEeqnarray}{rCL}\label{eq:discrete-time-IO_derive}
\mathsf{y}_k&=&\int_{-\infty}^{\infty}\mathsf{y}(\tau)p^{*}(\tau-(k-1)T_{S})\,\mathrm{d}\tau\\
&=&\frac{1}{\sqrt{T_{S}}}\int_{(k-1)T_{S}}^{kT_{S}}\mathsf{y}(\tau)\,\mathrm{d}\tau\\
&=&\frac{1}{\sqrt{T_{S}}}\int_{(k-1)T_{S}}^{kT_{S}}\hspace{-2mm}\left(\sum_{m=-M}^{M}\mathsf{s}_{m}e^{j2\pi\frac{m}{T_{S}N}\tau} \frac{\sqrt{\rho}\ \mathsf{x}_k}{\sqrt{T_{S}}}%\right.%\nonumber\\&&\rule{30mm}{0mm} 
%\left. \vphantom{\sum_{m=-M}^{M}\mathsf{s}_{m}}
+\mathsf{w}(\tau) \right)\,\mathrm{d}\tau \nonumber\\
\\
&=&\sqrt{\rho}\ \mathsf{x}_{k}\sum_{m=-M}^{M}\mathsf{s}_{m}e^{j2\pi \frac{m(k-1/2)}{N}}\sinc\lefto(\frac{m}{N}\right)+\mathsf{w}_{k}\label{eq:discrete-time-IO_derive3}
\end{IEEEeqnarray}
where $k=1,\dots,N$ with $N=T/T_{S}$ being the number of  symbols transmitted within each block and
\begin{IEEEeqnarray}{rCL}
  %\IEEEeqnarraymulticol{3}{l}{...}
  % a & = & b +c
\mathsf{w}_{k}=\frac{1}{\sqrt{T_{S}}}\int_{(k-1)T_{S}}^{k T_{S}}\mathsf{w}(\tau)\,\mathrm{d}\tau\,.
\end{IEEEeqnarray}
The additive noise random variables $\{\mathsf{w}_{k}\}$ are \iid zero-mean proper complex Gaussian.
To keep the notation simple and without loss of generality, we assume that the input-output relation is normalized and that the $\{\mathsf{w}_{k}\}$ have unit variance.
Then $\rho$ in~\eqref{eq:discrete-time-IO_derive3} can be thought of as the SNR.
Setting
\begin{IEEEeqnarray}{rCL}
  \mathsf{h}_k=\sum_{m=-M}^{M}\mathsf{s}_m e^{j2\pi \frac{m(k-1/2)}{N}}\sinc\lefto(\frac{m}{N}\right)
\end{IEEEeqnarray}
we rewrite~\eqref{eq:discrete-time-IO_derive3} in the following more compact form:
\begin{IEEEeqnarray}{rCL}\label{eq:discrete-time-IO}
  \mathsf{y}_k=\sqrt{\snr} \ \mathsf{h}_k \mathsf{x}_k + \mathsf{w}_k, \quad k=1,\dots,N.
\end{IEEEeqnarray}

Because of the block-memoryless assumption, the capacity $C(\rho)$ of the discrete-time channel~\eqref{eq:discrete-time-IO} is given by
\begin{IEEEeqnarray}{rCL}\label{eq:capacity}
    C(\rho)=\frac{1}{N} \sup I(\bm{\mathsf{x}};\bm{\mathsf{y}})
\end{IEEEeqnarray}
where $\bm{\mathsf{x}}=[\mathsf{x}_{1}\, \dots \,\mathsf{x}_{N}]^{T}$, $\bm{\mathsf{y}}=[\mathsf{y}_{1}\,\dots \,\mathsf{y}_{N}]^{T}$
and the supremum is over all probability measures on $\bm{\mathsf{x}}$ that satisfy the average-power constraint~$\sum_{k=1}^{N}\Ex{}{|\mathsf{x}_{k}|^2}\leq N$.
Note that to approach $C(\rho)$ one has to code over sufficiently many independent fading blocks.
No closed-form expressions for $C(\rho)$ are known for the case $N>1$.
Let the capacity pre-log $\chi$ be defined as
  \begin{IEEEeqnarray}{rCL}\label{eq:prelog1}
    \chi=\lim_{\snr\to\infty}\frac{C(\snr)}{\log \snr}.
  \end{IEEEeqnarray} 
It follows from~\cite[Th.~1]{liang04-09a} that 
\begin{IEEEeqnarray}{rCL}\label{eq:lv04_pre_log}
    \chi=1-\frac{Q}{N}
\end{IEEEeqnarray}
provided that $Q<N$.
The intuition behind this result is as follows~\cite{durisi11-08a}: $Q$ out of the $N$ available symbols  per block need to be sacrificed to learn the channels. 
This can be done for example by transmitting $Q$ pilot symbols per block.
The remaining $N-Q$ symbols can be used to communicate information. 
Hence, the capacity pre-log, which can be thought of as the number of ``dimensions'' per channel use available for communication, is $(N-Q)/N=1-Q/N$.
The ratio $Q/N$ corresponds to the ratio of the bandwidth $2\nu_{\textrm{max}}$ of the fading process to the symbol rate $1/T_{S}$:
\begin{IEEEeqnarray}{rCL}
\frac{Q}{N}&\approx& 2\nu_{\textrm{max}}T_{S}.
\end{IEEEeqnarray}

Note that the capacity pre-log~\eqref{eq:lv04_pre_log} is a lower bound on the capacity pre-log of the underlying continuous-time channel~\eqref{eq:continuous-time-IO} because~\eqref{eq:lv04_pre_log} is obtained i) by constraining the input signal to be of the form~\eqref{eq:TxSignal} and ii) by using symbol matched filtering at the receiver side. 
Both choices may be suboptimal.

\section{Oversampling the Output Signal} 
\label{sec:oversampling_approach}

\subsection{The Oversampled Input-Output Relation} % (fold)
\label{sec:the_oversampled_input_output_relation}

% subsection the_oversampled_input_output_relation (end)
We show in this section that the symbol matched filter approach reviewed in Section~\ref{sec:symbol matched filtering approach} is  suboptimal. 
Specifically, we prove that by oversampling the continuous-time channel output  by a factor two, a pre-log as large as $1-{1}/{N}$ can be achieved. 

Our oversampled, discrete-time, input-output relation is obtained as follows.
The received signal $\mathsf{y}(t)$ is filtered using a rectangular pulse whose width is half the symbol time (i.e., half the width of the transmit pulse $p(t)$).
The resulting filtered output signal is then sampled at twice the symbol rate.
Let 
\begin{IEEEeqnarray}{rCL}
  %\IEEEeqnarraymulticol{3}{l}{...}
  % a & = & b +c
  \mathsf{w}_n=\sqrt{\frac{2}{T_S}} \int_{(n-1)T_S/2}^{nT_S/2} \mathsf{w}(\tau)  \,\mathrm{d}\tau.
\end{IEEEeqnarray}
Within a fading block, the resulting $2N$ output samples are given by
\begin{IEEEeqnarray}{rCL}
  %\IEEEeqnarraymulticol{3}{l}{...}
  % a & = & b +c
  \mathsf{y}_n&=&\int_{-\infty}^{\infty} \mathsf{y}(\tau)\sqrt{\frac{2}{T_{S}}}\rect\lefto(\frac{\tau-T_S/4-(n-1)T_S/2}{T_S/2}\right)\mathrm{d}\tau \nonumber\\ \\
  &=& \sqrt{\frac{2}{T_S}}\int_{(n-1)T_S/2}^{nT_S/2}\mathsf{y}(\tau) \,\mathrm{d}\tau\\
&=&\sqrt{\frac{2}{T_S}}\int_{(n-1)T_S/2}^{nT_S/2}\left(\sum_{m=-M}^{M}\mathsf{s}_{m}e^{j2\pi\frac{m}{T_{S}N}\tau}\right.\nonumber\\&&\qquad\left.\qquad\qquad\vphantom{\sum_{m=-M}^{M}}\times\sqrt{\rho}\ \mathsf{x}_{\ceil{n/2}} \frac{1}{\sqrt{T_{S}}}+\mathsf{w}(\tau)\right) \,\mathrm{d}\tau\\
  &=&\sqrt{\frac{\rho}{2}}\ \mathsf{x}_{\ceil{n/2}}\sum_{m=-M}^M \mathsf{s}_m e^{j\pi\frac{m}{N}\left(n-\frac{1}{2}\right)} \sinc\lefto(\frac{m}{2N}\right) + \mathsf{w}_n
  ,\nonumber\\&&\qquad\qquad\qquad\qquad\qquad\qquad \quad n=1,\dots,2N. \label{eq:derivationyn}
\end{IEEEeqnarray}
Introducing the shorthand notation
\begin{IEEEeqnarray}{rCL}
  %\IEEEeqnarraymulticol{3}{l}{...}
  % a & = & b +c
  p_m=\frac{1}{\sqrt{2}}\ e^{-j\pi\frac{m}{2N}}\sinc\lefto(\frac{m}{2N}\right) \sqrt{\frac{1}{T}S_{\mathsf{h}}\lefto(\frac{m}{T}\right)}
\end{IEEEeqnarray}
we can rewrite~\eqref{eq:derivationyn} as
\begin{IEEEeqnarray}{rCL}
  %\IEEEeqnarraymulticol{3}{l}{...}
  % a & = & b +c
  \mathsf{y}_n&=& \sqrt{\rho}\ \mathsf{x}_{\ceil{n/2}}\sum_{m=-M}^M p_m\hat{\mathsf{s}}_m e^{j\pi\frac{mn}{N}} + \mathsf{w}_n
  , \quad n=1,\dots,2N.\nonumber\\ \label{eq:derivationyn2}
\end{IEEEeqnarray}
It turns out convenient to distinguish between the output samples corresponding to even values of~$n$ and the ones corresponding to odd values of $n$,
and to group them in two $N$-dimensional vectors as follows:
\begin{IEEEeqnarray}{rCL}
  %\IEEEeqnarraymulticol{3}{l}{...}
  % a & = & b +c
  \bm{\mathsf{y}}\sub{o}&=&\tp{[\mathsf{y}_1\,\, \mathsf{y}_3\,\, \dots \,\, \mathsf{y}_{2N-1}]} \\
  \bm{\mathsf{y}}\sub{e}&=&\tp{[\mathsf{y}_2\,\, \mathsf{y}_4\,\, \dots \,\, \mathsf{y}_{2N}]}.
\end{IEEEeqnarray}
Here, the subscripts ``o'' and ``e'' stand for odd and even, respectively.
To conveniently express the input-output relation in vector-matrix form, we next introduce some definitions.
Let $\bm{\mathsf{x}}=\tp{[\mathsf{x}_1\, \dots \,\mathsf{x}_N]}$, $\hat{\bm{\mathsf{s}}}=\tp{[\hat{\mathsf{s}}_{-M}\,\dots\,\hat{\mathsf{s}}_{M}]}$, and let $\bm{\mathsf{w}}\sub{o}$ and $\bm{\mathsf{w}}\sub{e}$ contain the odd and even noise samples, respectively.
Furthermore, define the $Q$-dimensional vector $\vecp$ (recall that $Q=2M+1$) as follows
\begin{IEEEeqnarray}{rCL}
  %\IEEEeqnarraymulticol{3}{l}{...}
  % a & = & b +c
  \vecp=\tp{[p_{-M}\, \dots \, p_M]}.
\end{IEEEeqnarray}
Finally, let 
\begin{IEEEeqnarray}{rCL}
  %\IEEEeqnarraymulticol{3}{l}{...}
  % a & = & b +c
  q\sub{o}^{(k,m)}&=&e^{j \pi \frac{m}{N}(2k-1)}, \quad k=1,\dots,N,\, m=-M,\dots,M \nonumber\\ \\
  \vecq\sub{o}^{(k)}&=&[q\sub{o}^{(k,-M)}\,\dots \, q\sub{o}^{(k,M)}],  \quad k=1,\dots,N\\[1.5mm]
  \matQ\sub{o}&=&\left[ 
  \begin{array}{c}
  \vecq\sub{o}^{(1)} \\
  \vdots\\
  \vecq\sub{o}^{(N)} 
  \end{array}
  \right].
\end{IEEEeqnarray}
Similarly, let
\begin{IEEEeqnarray}{rCL}
  %\IEEEeqnarraymulticol{3}{l}{...}
  % a & = & b +c
  q\sub{e}^{(k,m)}&=&e^{j \pi \frac{m}{N}(2k)}, \quad k=1,\dots,N,\, m=-M,\dots,M \nonumber\\ \\
  \vecq\sub{e}^{(k)}&=&[q\sub{e}^{(k,-M)}\,\dots \, q\sub{e}^{(k,M)}],  \quad k=1,\dots,N\\[1.5mm]
  \matQ\sub{e}&=&\left[ 
  \begin{array}{c}
  \vecq\sub{e}^{(1)} \\
  \vdots\\
  \vecq\sub{e}^{(N)} 
  \end{array}
  \right].
\end{IEEEeqnarray}

Equipped with these definitions and using~\eqref{eq:derivationyn2}, we can express $\bm{\mathsf{y}}\sub{o}$ and $\bm{\mathsf{y}}\sub{e}$ as
\begin{IEEEeqnarray}{rCL}\label{eq:systemmodel}
  \bm{\mathsf{y}}\sub{o}&=& \sqrt{\snr}\diag\{\bm{\mathsf{x}}\}\matQ\sub{o}\diag\{\mathbf{p}\}\hat{\bm{\mathsf{s}}}+\bm{\mathsf{w}}\sub{o}\IEEEyesnumber \IEEEyessubnumber \\
  \bm{\mathsf{y}}\sub{e}&=& \sqrt{\snr}\diag\{\bm{\mathsf{x}}\}\matQ\sub{e}\diag\{\mathbf{p}\}\hat{\bm{\mathsf{s}}}+\bm{\mathsf{w}}\sub{e}.\IEEEyessubnumber
\end{IEEEeqnarray}

\subsection{Pre-log Analysis} % (fold)
\label{sec:pre_log_analysis}
The capacity of the oversampled discrete-time channel (\ref{eq:systemmodel}) is given by
\begin{IEEEeqnarray}{rCL}\label{eq:CapOVS}
C(\snr)&=&\frac{1}{N}\sup  I(\bm{\mathsf{x}};\bm{\mathsf{y}}\sub{o},\bm{\mathsf{y}}\sub{e})
\end{IEEEeqnarray}
where the supremum is taken over all input distributions that satisfy the average power constraint $\sum_{k=1}^{N}\Ex{}{|\mathsf{x}_{k}|^{2}}\leq N$. 
The capacity pre-log is defined as in~\eqref{eq:prelog1}.
Our main result is given in the following theorem.
\begin{thm}\label{MainTheorem}
The capacity pre-log of the channel~\eqref{eq:systemmodel} is lower-bounded as
\be\label{eq:prelim}
\chi\geq 1-\frac{1}{N}.
\ee
\end{thm}

\begin{IEEEproof}
Our proof is based on the method proposed in~\cite{morgenshtern13-07a} and subsequently simplified in~\cite{koliander13-10a}. 

For convenience, we introduce the following notation 
\be\label{eq:modelsimp}
\bm{\mathsf{y}}=
\begin{pmatrix}
  \bm{\mathsf{y}}\sub{o}\\
  \bm{\mathsf{y}}\sub{e}
  \end{pmatrix}
  = \sqrt{\snr}\, \bar{\bm{\mathsf{y}}}
    +\bm{\mathsf{w}}
\ee
where 
\be\label{eq:ybar}
 \bar{\bm{\mathsf{y}}} =\bm{\mathsf{B}}
\hat{\bm{\mathsf{s}}}
\ee
with
\begin{IEEEeqnarray}{rCL}
  %\IEEEeqnarraymulticol{3}{l}{...}
  % a & = & b +c
  \bm{\mathsf{B}}=\begin{pmatrix}
   \diag\{\bm{\mathsf{x}}\}\matQ\sub{o} \\
    \diag\{\bm{\mathsf{x}}\}\matQ\sub{e}
    \end{pmatrix} \diag\{\mathbf{p}\}
\end{IEEEeqnarray}
and $\bm{\mathsf{w}}=\tp{(\tp{\bm{\mathsf{w}}\sub{o}}\;\;\tp{\bm{\mathsf{w}}\sub{e}})}$.
Let $h(\cdot)$ denote differential entropy.
We will establish a lower bound on the mutual information
\be\label{eq:decomposemutinf}
I(\bm{\mathsf{x}};\bm{\mathsf{y}}) =h(\bm{\mathsf{y}})-h(\bm{\mathsf{y}}|\bm{\mathsf{x}})
\ee
for the specific choice $\bm{\mathsf{x}} \sim \mathcal{CN}(\veczero,\matidentity_{N})$,
by upper-bounding $h(\bm{\mathsf{y}}|\bm{\mathsf{x}})$ and lower-bounding $h(\bm{\mathsf{y}})$.

\newcounter{tempequationcounter}
\begin{figure*}[!b]
  \vspace*{4pt}  
  \hrulefill
\normalsize
  \setcounter{tempequationcounter}{\value{equation}}
  \begin{IEEEeqnarray}{rCl}
\setcounter{equation}{60} 
\matJ_{\phi_{x_1}}(\hat{\vecs},[\vecx]^{}_{[2:N]}) &=& \begin{pmatrix}
\diag\{\vecx\}\matQ\sub{o}\diag\{\mathbf{p}\} & \begin{matrix}  \veczero_{1\times N-1}\\ \diag\{[\matQ\sub{o}]_{[2:N]}\diag\{\mathbf{p}\}\hat{\vecs}\}  \end{matrix} \\
\diag\{[\vecx]\sub{[1:Q-1]}\}[\matQ\sub{e}]_{[1:Q-1]}\diag\{\mathbf{p}\} & \begin{matrix}\veczero_{1\times N-1}\\
\begin{matrix}\diag\{[\matQ\sub{e}]_{[2:Q-1]}\diag\{\mathbf{p}\}\hat{\vecs}\} & \veczero_{Q-2\times N-Q+1}\end{matrix}\end{matrix}
\end{pmatrix}\label{eq:jac_phi}
  \end{IEEEeqnarray}
  \setcounter{equation}{\value{tempequationcounter}}
\end{figure*}

\subsubsection*{Upper bound on $h(\bm{\mathsf{y}}|\bm{\mathsf{x}})$}
Since $\hat{\bm{\mathsf{s}}}\distas \jpg\lefto(\veczero,\matidentity_{Q}\right)$ and $\bm{\mathsf{w}}\sim\mathcal{CN}(\veczero,\matidentity_{2N})$, we conclude that $\bm{\mathsf{y}}$ is conditionally Gaussian  given $\bm{\mathsf{x}}$, with 
conditional covariance matrix 
\begin{IEEEeqnarray}{rCL}
  %\IEEEeqnarraymulticol{3}{l}{...}
  % a & = & b +c
  \Ex{}{\bm{\mathsf{y}}\herm{\bm{\mathsf{y}}}\given \bm{\mathsf{x}}}= \rho \bm{\mathsf{B}}\herm{\bm{\mathsf{B}}}+\matidentity_{2N}.
\end{IEEEeqnarray}
Hence,~\cite[Th.~2]{NeeserMassey1993}
\begin{IEEEeqnarray}{rCL}
  %\IEEEeqnarraymulticol{3}{l}{...}
  % a & = & b +c
  h(\bm{\mathsf{y}}|\bm{\mathsf{x}}) 
  = \opE_{\bm{\mathsf{x}}}\big[\log\bigl((\pi e)^{2N}\det\bigl(\rho \bm{\mathsf{B}}\bm{\mathsf{B}}^{\operatorname{H}}+ \matidentity_{2N}\big) \bigr)\big].
\end{IEEEeqnarray}
We next use that 
$\det\bigl(\rho \bm{\mathsf{B}}\bm{\mathsf{B}}^{\operatorname{H}}+\matidentity_{2N}\bigr) 
= \det\bigl(\rho\bm{\mathsf{B}}^{\operatorname{H}}\bm{\mathsf{B}} + \matidentity_{Q}\big)$ (which follows from~\cite[Th.~1.3.20]{hojo85}).  
Furthermore, assuming without loss of generality  
that $\rho > 1$ (note that we are only interested in the asymptotic regime $\rho\rightarrow \infty$), we have
$\det\bigl(\rho \bm{\mathsf{B}}^{\operatorname{H}}\bm{\mathsf{B}} + \matidentity_{Q}\big) \leq \det\bigl(\rho \big(\bm{\mathsf{B}}^{\operatorname{H}}\bm{\mathsf{B}} + \matidentity_{Q}\big)\big) = \rho^{Q}\det\bigl( \bm{\mathsf{B}}^{\operatorname{H}}\bm{\mathsf{B}} + \matidentity_{Q}\big)$.
Thus,
\begin{align}
h(\bm{\mathsf{y}} |\bm{\mathsf{x}})& \leq \opE_{\bm{\mathsf{x}}}\big[\log\big((\pi e)^{2N}\rho^{Q} \det\bigl( \bm{\mathsf{B}}^{\operatorname{H}}\bm{\mathsf{B}} +\matidentity_{Q}\bigr) \big)\big] \notag \\[.5mm]
&=Q\log \rho + \opE_{\bm{\mathsf{x}}}\big[\log\det\bigl( \bm{\mathsf{B}}^{\operatorname{H}}\bm{\mathsf{B}}+ \matidentity_{Q}\bigr) \big]+ \mathcal{O}(1)\,.\label{eq:hygivenxbound}
\end{align}
By applying Jensen's inequality to the concave function $\log(\cdot)$, we obtain  
\begin{align}\label{eq:showconst}
\opE_{\bm{\mathsf{x}}}\big[\log\det\bigl(\bm{\mathsf{B}}^{\operatorname{H}}\bm{\mathsf{B}}+ \matidentity_{Q}\big)\big] & 
\leq \log \opE_{\bm{\mathsf{x}}}\big[\det\bigl( \bm{\mathsf{B}}^{\operatorname{H}}\bm{\mathsf{B}}+\matidentity_{Q}\big)\big].
\end{align}
The determinant $\det\bigl( \bm{\mathsf{B}}^{\operatorname{H}}\bm{\mathsf{B}}+ \matidentity_{Q}\big)$ on the RHS of~\eqref{eq:showconst} is a polynomial in the entries of $\bm{\mathsf{x}}$ and $\bm{\mathsf{x}}^{\operatorname{H}}$. 
Since $\bm{\mathsf{x}} \sim \mathcal{CN}(\veczero,\matidentity_{N})$, all moments of $\bm{\mathsf{x}}$ are finite; hence, also the expectation $\opE_{\bm{\mathsf{x}}}\big[\det\bigl( \bm{\mathsf{B}}^{\operatorname{H}}\bm{\mathsf{B}}+ \matidentity_{Q}\big)\big]$ is finite. 
Thus, 
the right-hand side of~\eqref{eq:showconst} is a finite constant that does not depend on~$\rho$.
Hence,~\eqref{eq:hygivenxbound} together with~\eqref{eq:showconst} implies that
\be
h(\bm{\mathsf{y}} | \bm{\mathsf{x}}) \leq Q \log \rho + \mathcal{O}(1)\,. \label{eq:boundhygivenx}
\ee

\subsubsection*{Lower bound on $h(\bm{\mathsf{y}})$}
Define $\setI=[1:N+Q-1]$ and $\setJ=[N+Q:2N]$.
We can now lower-bound $h(\bm{\mathsf{y}})$ as follows:
\begin{align}
h(\bm{\mathsf{y}}) &=h([\bm{\mathsf{y}}]^{}_{\setI}, [\bm{\mathsf{y}}]^{}_{\setJ})\\
&=h([\bm{\mathsf{y}}]^{}_{\setI})+h\big([\bm{\mathsf{y}}]^{}_{\setJ} \big|[\bm{\mathsf{y}}]^{}_{\setI}\big)\label{eq:step_a}\\
&\geq h\bigl(\sqrt{\rho}[\bar{\bm{\mathsf{y}}}]^{}_{\setI}+[\bm{\mathsf{w}}]^{}_{\setI} \big|[\bm{\mathsf{w}}]^{}_{\setI}\bigr)+h\big([\bm{\mathsf{y}}]^{}_{\setJ} \big|\hat{\bm{\mathsf{s}}},\bm{\mathsf{x}},[\bm{\mathsf{y}}]^{}_{\setI}\big)\label{eq:step_b}\\
& = h\bigl(\sqrt{\rho}[\bar{\bm{\mathsf{y}}}]^{}_{\setI}\bigr)+\mathcal{O}(1)\label{eq:step_c}\\
%& = \log\bigl(\rho^{(N+Q-1)}\bigr) + h([\bar{\bm{\mathsf{y}}}]^{}_{\setI})+\mathcal{O}(1) \label{eq:step_d}\\
& = (N+Q-1) \log \rho 
 + h ([\bar{\bm{\mathsf{y}}}]^{}_{\setI})+\mathcal{O}(1)\,. \label{eq:boundhy}
\end{align}
Here,~\eqref{eq:step_a} follows from the chain rule for differential entropy~\cite[Th.~8.6.2]{Cover91}, in~\eqref{eq:step_b} we use~\eqref{eq:modelsimp} and the fact that conditioning reduces differential entropy,~\eqref{eq:step_c} holds since $h\big([\bm{\mathsf{y}}]^{}_{\setJ} \big|\hat{\bm{\mathsf{s}}},\bm{\mathsf{x}},[\bm{\mathsf{y}}]^{}_{\setI}\big)=h\big([\bm{\mathsf{w}}]^{}_{\setJ}\big)$ is a finite constant that is independent of $\snr$, and~\eqref{eq:boundhy} follows by the transformation property of differential entropy reported in~\cite[Eq.~(8.71)]{Cover91}.
Using~\eqref{eq:boundhygivenx} and~\eqref{eq:boundhy} in~\eqref{eq:decomposemutinf},
 we obtain
\begin{align}
I(\bm{\mathsf{x}} ;\bm{\mathsf{y}}) 
& = (N-1)\log \rho + h ([\bar{\bm{\mathsf{y}}}]^{}_{\setI})  + \mathcal{O}(1).
\label{EQmutual}
\end{align}
If we now divide the RHS of~\eqref{EQmutual} by $N\log\snr$ and then take the limit $\snr\to\infty$, we obtain the desired pre-log lower bound~\eqref{eq:prelim} 
provided that we are able to prove that $h ([\bar{\bm{\mathsf{y}}}]^{}_{\setI})  > -\infty$.

To conclude the proof, we will next show that indeed $h ([\bar{\bm{\mathsf{y}}}]^{}_{\setI})  > -\infty$.
Because conditioning reduces entropy, we have that
\be \label{eq:boundhycond}
h ([\bar{\bm{\mathsf{y}}}]^{}_{\setI})\geq h ([\bar{\bm{\mathsf{y}}}]^{}_{\setI}| \mathsf{x}_1)\,.
\ee
Coarsely speaking, conditioning  on $\mathsf{x}_1$ in~\eqref{eq:boundhycond} corresponds to transmitting one pilot symbol per fading block.
We will use the parametrized mappings
\begin{IEEEeqnarray}{rCL}
\phi_{x_1}\colon \complexset^{N+Q-1}\to \complexset^{N+Q-1};\quad (\hat{\vecs},[\vecx]^{}_{[2:N]})\mapsto [\bar{\vecy}]^{}_{\setI}
\end{IEEEeqnarray}
to establish a connection between $h ([\bar{\bm{\mathsf{y}}}]^{}_{\setI}| \mathsf{x}_1)$ and $h(\hat{\bm{\mathsf{s}}},[\bm{\mathsf{x}}]_{[2:N]})$, which is finite by construction. Let $\matJ_{\phi_{x_1}}(\hat{\vecs},[\vecx]_{[2:N]})$ be the Jacobian of the mapping $\phi_{x_1}$ given by~\eqref{eq:jac_phi}\addtocounter{equation}{1} at the bottom of the page. 
%\begin{IEEEeqnarray}{rCL}\label{eq:jac_phi}
%&&\matJ_{\phi_{x_1}}([\vecx]_{[2:N]},\hat{\vecs})\nonumber\\
%&&\ =\!\begin{pmatrix}
%\diag\{\vecx\}\matQ\sub{o}\diag\{\mathbf{p}\} & \begin{matrix}  \veczero_{1\times N-1}\\ \diag\{[\matQ\sub{o}]_{[2:N]}\diag\{\mathbf{p}\}\hat{\vecs}\}  \end{matrix} \\
%\diag\{[\vecx]\sub{[1:Q-1]}\}[\matQ\sub{e}]_{[1:Q-1]}\diag\{\mathbf{p}\} & \begin{matrix}\veczero_{1\times N-1}\\
%\begin{matrix}\diag\{[\matQ\sub{e}]_{[2:Q-1]}\diag\{\mathbf{p}\}\hat{\vecs}\} & \veczero_{Q-2\times N-Q+1}\end{matrix}\end{matrix}
%\end{pmatrix}\!\! .\hspace{0.6cm}
%\end{IEEEeqnarray}
%
Note that in~\eqref{eq:jac_phi} we did not take the derivative with respect to $x_1$ because it is treated as a parameter. 
By the definition of $\bar{\vecy}$ (see~\eqref{eq:ybar}),  $\phi_{x_1}$ is a vector-valued polynomial mapping. 
Thus, by~\cite[Lem.~7]{koliander13-10a} the function $\phi_{x_1}$ is an $m$-to-one mapping on the set $\widetilde{\setM}\triangleq \set{(\hat{\vecs},[\vecx]_{[2:N]}): \lvert\matJ_{\phi_{x_1}}(\hat{\vecs},[\vecx]_{[2:N]})\rvert\neq 0}$, i.e., there exists a finite $m\in \mathbb{N}$ such that the intersection of the inverse image $\phi_{x_1}^{-1}(\{[\bar{\vecy}]^{}_{\setI}\})=\set{(\hat{\vecs},[\vecx]_{[2:N]}): \phi_{x_1}(\hat{\vecs},[\vecx]_{[2:N]})=[\bar{\vecy}]^{}_{\setI}}$ and the set $\widetilde{\setM}$ contains at most $m$ elements. 
The value taken by $m$ depends only on the degree of the polynomial $\phi_{x_1}$ and not on the specific choice of $x_1\in \complexset$ and of $[\bar{\vecy}]^{}_{\setI}\in \complexset^{N+Q-1}$.
%We denote this number by $m$. 
%Here, the Jacobian $\matJ_{\phi_{x_1}}([\vecx]_{[2:N]},\hat{\vecs})$ of the mapping $\phi_{x_1}$ is given by
%Furthermore, the number $m$ does only depend on the degree of the polynomial and not on the specific choice of $x_1$.
%
%We will next show that  $\widetilde{\setM}\neq\emptyset$.
Since a polynomial (in our case, $\det\bigl(\matJ_{\phi_{x_1}}(\hat{\vecs},[\vecx]_{[2:N]})\bigr)$) either vanishes identically (which would imply $\widetilde{\setM}=\emptyset$) or vanishes on a set of measure zero (which would imply $\widetilde{\setM}\neq \emptyset$), showing that $\widetilde{\setM}\neq \emptyset$ is sufficient to conclude that $\det\bigl(\matJ_{\phi_{x_1}}(\hat{\vecs},[\vecx]_{[2:N]})\bigr)$ is nonzero almost everywhere.\footnote{Recall that in our notation $\abs{\matA}=\abs{\det(\matA)}$.} In the following lemma, whose proof can be found in Appendix~\ref{app:lemma}, we show that $\widetilde{\setM}\neq\emptyset$. 
\begin{lem}\label{lem:nonzero_element}
For almost all $x_1$ there exists a pair $(\hat{\vecs},[\vecx]_{[2:N]})$ for which $\lvert\matJ_{\phi_{x_1}}(\hat{\vecs},[\vecx]_{[2:N]})\rvert\neq 0$.
\end{lem}

Lemma~\ref{lem:nonzero_element} implies that for almost all $x_1$ the function $\phi_{x_1}$ is $m$-to-one almost everywhere.
Thus, we can now use the transformation rule for differential entropy under a finite-to-one mapping established in~\cite[Lem.~8]{koliander13-10a}
\ba\label{eq:applylemma8}
&h\big(\phi_{x_1}(\hat{\bm{\mathsf{s}}},[\bm{\mathsf{x}}]^{}_{[2:N]})\big)  \geq h(\hat{\bm{\mathsf{s}}},[\bm{\mathsf{x}}]^{}_{[2:N]}) - \log m \notag \\
 & \rule{3mm}{0mm} + \int_{\complexset^{Q+N-1}} \! f_{\hat{\bm{\mathsf{s}}},[\bm{\mathsf{x}}]^{}_{[2:N]}}(\hat{\vecs},[\vecx]^{}_{[2:N]})\nonumber\\
& \rule{18mm}{0mm} \times \log \big(\big\lvert\matJ_{\phi_{x_1}}\!(\hat{\vecs},[\vecx]^{}_{[2:N]})\big\rvert^2\big)  \mathrm{d}(\hat{\vecs},[\vecx]^{}_{[2:N]})\,.
\ea
Because of $[\bar{\vecy}]^{}_{\setI}=\phi_{x_1}(\hat{\vecs},[\vecx]^{}_{[2:N]})$, we have $h\big([\bar{\bm{\mathsf{y}}}]^{}_{\setI} \big| \mathsf{x}_1\!=\!x_1\big)= h\big(\phi_{x_1}(\hat{\bm{\mathsf{s}}},[\bm{\mathsf{x}}]^{}_{[2:N]})\big)$.
Thus,~\eqref{eq:applylemma8} entails
\begin{align}
&h\big([\bar{\bm{\mathsf{y}}}]^{}_{\setI} \big| \mathsf{x}_1\big) \,  \geq\, h(\hat{\bm{\mathsf{s}}},[\bm{\mathsf{x}}]^{}_{[2:N]}) - \log m \notag \\
&\quad  + \opE_{\mathsf{x}_1} \left[\rule{0cm}{8mm}\right.\int_{\complexset^{Q+N-1}} \! f_{\hat{\bm{\mathsf{s}}},[\bm{\mathsf{x}}]^{}_{[2:N]}}(\hat{\vecs},[\vecx]^{}_{[2:N]}) \nonumber\\
&\rule{20mm}{0mm} \times\log \bigl(\big\lvert\matJ_{\phi_{x_1}}\!(\hat{\vecs},[\vecx]^{}_{[2:N]})\big\rvert^2\big)  \mathrm{d}(\hat{\vecs},[\vecx]^{}_{[2:N]}) \left.\rule{0cm}{8mm}\right] . \label{eq:boundhybar}
\end{align}

\begin{figure*}[!b]
  \vspace*{4pt}  
  \hrulefill
\normalsize
  \setcounter{tempequationcounter}{\value{equation}}
\begin{IEEEeqnarray}{rCL}
\setcounter{equation}{67} 
\matJ_{\phi_{x_1}}(\hat{\vecs},[\vecx]^{}_{[2:N]})=\begin{pmatrix}
\diag\{\vecx\}\matQ\sub{o}\diag\{\mathbf{p}\} & \begin{matrix}  \veczero_{Q-1\times Q-2} & \veczero_{Q-1\times N-Q+1}\\ 
\matzero_{N-Q+1\times Q-2} &
\matD_2  \end{matrix} \\
\diag\{[\vecx]\sub{[1:Q-1]}\}[\matQ\sub{e}]_{[1:Q-1]}\diag\{\mathbf{p}\} & \begin{matrix}\veczero_{1\times Q-2} & \veczero_{1\times N-Q+1} \\
\matD_1 & \veczero_{Q-2\times N-Q+1}\end{matrix}
\end{pmatrix}\label{eq:JacobianChoice}
\end{IEEEeqnarray}
    \setcounter{equation}{\value{tempequationcounter}}
\end{figure*}

We now show that the RHS of~\eqref{eq:boundhybar} is lower-bounded by a finite constant.
The term $h(\hat{\bm{\mathsf{s}}},[\bm{\mathsf{x}}]^{}_{[2:N]})$ is the differential entropy of a standard multivariate proper complex Gaussian random vector and thus a finite constant. 
Hence, it remains to characterize
\begin{align}
&\hspace{-1mm}\opE_{\mathsf{x}_1} \!\left[\rule{0cm}{8mm}\right.\int_{\complexset^{Q+N-1}} \! f_{\hat{\bm{\mathsf{s}}},[\bm{\mathsf{x}}]^{}_{[2:N]}}(\hat{\vecs},[\vecx]^{}_{[2:N]}) \nonumber\\
&\rule{14mm}{0mm}\times\log \big(\big\lvert\matJ_{\phi_{x_1}}\!(\hat{\vecs},[\vecx]^{}_{[2:N]})\big\rvert^2\big) 
   \mathrm{d}(\hat{\vecs},[\vecx]^{}_{[2:N]})\left.\rule{0cm}{8mm}\right] \\
&=
\int_{\complexset}\int_{\complexset^{Q+N-1}} \! f_{\mathsf{x}_1}(x_1)\,f_{\hat{\bm{\mathsf{s}}},[\bm{\mathsf{x}}]^{}_{[2:N]}}(\hat{\vecs},[\vecx]^{}_{[2:N]})\nonumber\\
&\rule{14mm}{0mm}\times \log \big(\big\lvert\matJ_{\phi_{x_1}}\!(\hat{\vecs},[\vecx]^{}_{[2:N]})\big\rvert^2\big)  \mathrm{d}(\hat{\vecs},[\vecx]^{}_{[2:N]})\, \mathrm{d}x_1\\
&=
\int_{\complexset^{Q+N}} \! f_{\hat{\bm{\mathsf{s}}},\bm{\mathsf{x}}}(\hat{\vecs},\vecx) \log \bigl(\big\lvert\matJ_{\phi_{x_1}}\!(\hat{\vecs},[\vecx]^{}_{[2:N]})\big\rvert^2\big)  \mathrm{d}(\hat{\vecs},\vecx)
\label{eq:finlogdet}
\end{align}
where~\eqref{eq:finlogdet} holds because $(\hat{\bm{\mathsf{s}}},[\bm{\mathsf{x}}]^{}_{[2:N]})$ and $\mathsf{x}_1$ are independent.
We use the following result from~\cite[Lem.~9]{koliander13-10a} to show that the RHS of~\eqref{eq:finlogdet} is bounded away from minus infinity.
\begin{lem}\label{LEMboundanalytic}
Let $f$ be an analytic function on $\complexset^n$ that is not identically zero. Then
\be\label{eq:expec}
I_1 \triangleq
\int_{\complexset^n} \!\exp(-\vecnorm{\vecxi}^2)\log(\abs{f(\vecxi)})\,\mathrm{d}\vecxi > -\infty \,.
\vspace{1.5mm}
\ee
\end{lem}

Since the determinant of $\matJ_{\phi_{x_1}}\!({\hat{\vecs}},[\vecx]^{}_{[2:N]})$ is a complex polynomial that is nonzero a.e., it is an analytic function that is not identically zero. 
Furthermore, $f_{\hat{\bm{\mathsf{s}}},\bm{\mathsf{x}}}$ is the  probability density function of a standard multivariate Gaussian random vector.
Hence, by Lemma~\ref{LEMboundanalytic}, the RHS in~\eqref{eq:finlogdet} is bounded away from minus infinity. 
Thus, using~\eqref{eq:boundhybar}, we conclude that $h\big([\bar{\bm{\mathsf{y}}}]^{}_{\setI}\big|\mathsf{x}_1\big)>-\infty$. 
Together with~\eqref{eq:boundhycond}, this implies  $h([\bar{\bm{\mathsf{y}}}]^{}_{\setI})>-\infty$. 
This concludes the proof. 
\end{IEEEproof}

Note that the assumption that $\hat{\bm{\mathsf{s}}}$ is complex-Gaussian distributed (Rayleigh fading) can be partially relaxed in the proof of Theorem~\ref{MainTheorem}. 
Indeed, Theorem~\ref{MainTheorem} holds for any circularly symmetric fading distribution $f_{\hat{\bm{\mathsf{s}}}}$ that decays fast enough\footnote{A detailed analysis of~\cite[Appendix~D]{morgenshtern13-07a} shows that the decay has to be such that $\opE_{\hat{\bm{\mathsf{s}}},\bm{\mathsf{x}}}\Big[\log\big(\vecnorm{(\hat{\bm{\mathsf{s}}}^{T}\,\bm{\mathsf{x}}^{T})}\big) \vecnorm{(\hat{\bm{\mathsf{s}}}^{T}\, \bm{\mathsf{x}}^{T})}^{2(Q+N)+1}\Big]$ is finite.}
and for which $h(\hat{\bm{\mathsf{s}}})>-\infty$.
%However, it is not clear what assuptions on the continuous time fading process would result in a discrete time model with fading statistics.

% section pre_log_analysis (end)

% subsection oversampling_approach (end)

% section system (end)

\section{Conclusions}\label{Sect:Conclusion}
We have shown that the capacity pre-log of a continuous-time, time-selective, Rayleigh block-fading channel is lower-bounded by $1-1/N$. This pre-log, which can be achieved by sampling the channel output at twice the symbol rate, is independent of the rank $Q$  of the covariance matrix characterizing the temporal correlation of the fading inside each fading block. 
In contrast, the standard symbol matched filtering approach, which entails sampling at symbol rate, leads to the looser lower bound $1-Q/N$.

As already discussed in the introduction, symbol rate sampling does not yield in general a sufficient statistics for the detection of the transmitted symbols from the output samples.
This is due to the bandwidth expansion resulting from the multiplication of the channel input process by the fading process. 

Coarsely speaking, oversampling yields an increase of the dimension of the output space spanned by the received samples.
The resulting additional dimensions   can be used to acquire knowledge about the fading channel at the receiver.
Indeed, as illustrated in Section~\ref{sec:pre_log_analysis} (see~\eqref{eq:boundhycond}) one pilot symbol per fading block is sufficient for the case of oversampling, whereas $Q$ pilot symbols are required for the case of symbol matched filtering. This explains the pre-log increase resulting from oversampling.
Unfortunately, the processing needed to acquire this additional channel knowledge is nonlinear.
In contrast, standard minimum mean-square estimation of the fading channel based on the pilot symbols is pre-log optimal for the case of symbol matched filtering~\cite{zheng02-02a}.
This nonlinear processing is the reason why the proof of our main result is technical in some parts.
A phenomenon similar to the one just described has been recently observed in the context of multiple-antenna communications, where increasing the number of receive antennas yield a pre-log increase for block-correlated fading channels, even when the transmitter has a single antenna~\cite{morgenshtern13-07a,koliander13-10a}.

\appendices

\section{Proof of Lemma~\ref{lem:nonzero_element}} % (fold)
\label{app:lemma}
\begin{IEEEproof}
It is convenient to choose $\hat{\vecs}$ so that $\diag\{\mathbf{p}\}\hat{\vecs}$ is nonzero and orthogonal to the rows of the matrix $[\matQ\sub{o}]_{[1:Q-1]}$. 
Note that the elements of $\mathbf{p}$ are nonzero by construction. 
Moreover, the matrix $(\tp{\matQ\sub{o}} \; \tp{\matQ\sub{e}})$ is \emph{full spark} \cite[Def.~1]{alcami12}, i.e., every set of $Q$ columns of $(\tp{\matQ\sub{o}} \; \tp{\matQ\sub{e}})$ is linearly independent, because it is a Vandermonde matrix with nonequal columns (see~\cite[Lem.~2]{alcami12}). 
Thus, our choice of $\hat{\vecs}$ yields $[\matQ\sub{o}]_{\{k\}} \diag\{\mathbf{p}\}\hat{\vecs}\neq 0$ for $k\in [Q:N]$ and $[\matQ\sub{e}]_{\{k\}}\diag\{\mathbf{p}\} \hat{\vecs}\neq 0$ for $k\in [1:N]$. 
To simplify notation, we set $\matD_1=\diag\{[\matQ\sub{e}]_{[2:Q-1]}\diag\{\mathbf{p}\}\hat{\vecs}\}$ and $\matD_2=\diag\{[\matQ\sub{o}]_{[Q:N]}\diag\{\mathbf{p}\}\hat{\vecs}\}$. 
Note that $\matD_1$ and $\matD_2$  are diagonal matrices with nonzero diagonal entries. 
We can rewrite $\matJ_{\phi_{x_1}}(\hat{\vecs},[\vecx]_{[2:N]})$ as shown in (\ref{eq:JacobianChoice})\addtocounter{equation}{1} at the bottom of the page.
%
%\begin{IEEEeqnarray}{rCL}
%\matJ_{\phi_{x_1}}([\vecx]_{[2:N]},\hat{\vecs})=\begin{pmatrix}
%\diag\{\vecx\}\matQ\sub{o}\diag\{\mathbf{p}\} & \begin{matrix}  \veczero_{Q-1\times Q-2} & \veczero_{Q-1\times N-Q+1}\\ 
%\matzero_{N-Q+1\times Q-2} &
%\matD_2  \end{matrix} \\
%\diag\{[\vecx]\sub{[1:Q-1]}\}[\matQ\sub{e}]_{[1:Q-1]}\diag\{\mathbf{p}\} & \begin{matrix}\veczero_{1\times Q-2} & \veczero_{1\times N-Q+1} \\
%\matD_1 & \veczero_{Q-2\times N-Q+1}\end{matrix}
%\end{pmatrix}\!.\hspace{0.6cm}\label{eq:JacobianChoice}
%\end{IEEEeqnarray}
%
The determinant of $\matJ_{\phi_{x_1}}(\hat{\vecs},[\vecx]_{[2:N]})$ can now be factorized as follows
\begin{align}
&\hspace{-3mm}\abs{\matJ_{\phi_{x_1}}(\hat{\vecs},[\vecx]_{[2:N]})} \nonumber\\
&=\left\lvert\rule{0cm}{5.8mm}\right.{\underbrace{\begin{pmatrix}x_1[\matQ\sub{e}]_{\{1\}}\\
\diag\{[\vecx]\sub{[1:Q-1]}\}[\matQ\sub{o}]\sub{[1:Q-1]}\end{pmatrix}\diag\{\mathbf{p}\}
}_{=\mathbf{A}}}\left.\rule{0cm}{5.8mm}\right\rvert \abs{\matD_2} \abs{\matD_1}\,.
\end{align}
Choosing $x_1$ so that  $x_1\neq 0$ (recall that we need to establish that $\abs{\matJ_{\phi_{x_1}}(\hat{\vecs},[\vecx]_{[2:N]})}\neq 0$ only for almost all $x_1$) and choosing all other entries of $\vecx$ also nonzero yields $\abs{\mathbf{A}} \neq 0$ as the matrix $\mathbf{A}$ is the product of the nonsingular matrices $\diag\{\mathbf{p}\}$ and $\begin{pmatrix}x_1[\matQ\sub{e}]_{\{1\}}\\
\diag\{[\vecx]\sub{[1:Q-1]}\}[\matQ\sub{o}]\sub{[1:Q-1]}\end{pmatrix}$ (recall that $(\tp{\matQ\sub{o}} \; \tp{\matQ\sub{e}})$ is full spark). 
Furthermore, we have that $\abs{\matD_2}\neq 0$ and $\abs{\matD_1}\neq 0$.
Hence, we conclude that  $\abs{\matJ_{\phi_{x_1}}(\hat{\vecs},[\vecx]_{[2:N]})}\neq 0$.
\end{IEEEproof}

% section ciao2 (end)

\enlargethispage{-10cm}

%%%%%%%%%%%%%%%%%%%%%%%%%%%%
\bibliographystyle{IEEEtran}
\bibliography{IEEEabrv,publishers,confs-jrnls,giubib,KoGbib,MDbib}

\begin{thebibliography}{10}
\providecommand{\url}[1]{#1}
\csname url@samestyle\endcsname
\providecommand{\newblock}{\relax}
\providecommand{\bibinfo}[2]{#2}
\providecommand{\BIBentrySTDinterwordspacing}{\spaceskip=0pt\relax}
\providecommand{\BIBentryALTinterwordstretchfactor}{4}
\providecommand{\BIBentryALTinterwordspacing}{\spaceskip=\fontdimen2\font plus
\BIBentryALTinterwordstretchfactor\fontdimen3\font minus
  \fontdimen4\font\relax}
\providecommand{\BIBforeignlanguage}[2]{{%
\expandafter\ifx\csname l@#1\endcsname\relax
\typeout{** WARNING: IEEEtran.bst: No hyphenation pattern has been}%
\typeout{** loaded for the language `#1'. Using the pattern for}%
\typeout{** the default language instead.}%
\else
\language=\csname l@#1\endcsname
\fi
#2}}
\providecommand{\BIBdecl}{\relax}
\BIBdecl

\bibitem{marzetta99-01a}
T.~L. Marzetta and B.~M. Hochwald, ``Capacity of a mobile multiple-antenna
  communication link in {Rayleigh} flat fading,'' \emph{{IEEE} Trans. Inf.
  Theory}, vol.~45, no.~1, pp. 139--157, Jan. 1999.

\bibitem{hochwald00-03a}
B.~M. Hochwald and T.~L. Marzetta, ``Unitary space--time modulation for
  multiple-antenna communications in {Rayleigh} flat fading,'' \emph{{IEEE}
  Trans. Inf. Theory}, vol.~46, no.~2, pp. 543--564, Mar. 2000.

\bibitem{zheng02-02a}
L.~Zheng and D.~N.~C. Tse, ``Communication on the {Grassmann} manifold: A
  geometric approach to the noncoherent multiple-antenna channel,''
  \emph{{IEEE} Trans. Inf. Theory}, vol.~48, no.~2, pp. 359--383, Feb. 2002.

\bibitem{yang13-02a}
W.~Yang, G.~Durisi, and E.~Riegler, ``On the capacity of large-{MIMO}
  block-fading channels,'' \emph{{IEEE} J. Sel. Areas Commun.}, vol.~31, no.~2,
  pp. 117--132, Feb. 2013.

\bibitem{lapidoth03-10a}
A.~Lapidoth and S.~M. Moser, ``Capacity bounds via duality with applications to
  multiple-antenna systems on flat-fading channels,'' \emph{{IEEE} Trans. Inf.
  Theory}, vol.~49, no.~10, pp. 2426--2467, Oct. 2003.

\bibitem{lapidoth06-02a}
------, ``The fading number of single-input multiple-output fading channels
  with memory,'' \emph{{IEEE} Trans. Inf. Theory}, vol.~52, no.~2, pp.
  437--453, Feb. 2006.

\bibitem{moser09-06a}
S.~M. Moser, ``{The fading number of multiple-input multiple-output fading
  channels with memory},'' \emph{{IEEE} Trans. Inf. Theory}, vol.~55, no.~6,
  pp. 2716--2755, Jun. 2009.

\bibitem{koch09-05a}
T.~Koch, \emph{On Heating Up and Fading in Communication Channels}, ser.
  Information Theory and its Applications, A.~Lapidoth, Ed.\hskip 1em plus
  0.5em minus 0.4em\relax Konstanz, Germany: Hartung-Gorre Verlag, May 2009,
  vol.~5.

\bibitem{lapidoth05-07a}
A.~Lapidoth, ``On the asymptotic capacity of stationary {Gaussian} fading
  channels,'' \emph{{IEEE} Trans. Inf. Theory}, vol.~51, no.~2, pp. 437--446,
  Feb. 2005.

\bibitem{durisi12-10a}
G.~Durisi, V.~I. Morgenshtern, and H.~B{\"o}lcskei, ``On the sensitivity of
  continuous-time noncoherent fading channel capacity,'' \emph{{IEEE} Trans.
  Inf. Theory}, vol.~58, no.~10, pp. 6372--6391, Oct. 2012.

\bibitem{liang04-09a}
Y.~Liang and V.~V. Veeravalli, ``Capacity of noncoherent time-selective
  {Rayleigh}-fading channels,'' \emph{{IEEE} Trans. Inf. Theory}, vol.~50,
  no.~12, pp. 3095--3110, Dec. 2004.

\bibitem{morgenshtern13-07a}
V.~I. Morgenshtern, E.~Riegler, W.~Yang, G.~Durisi, S.~Lin, B.~Sturmfels, and
  H.~B\"{o}lcskei, ``Capacity pre-log of noncoherent {SIMO} channels via
  {Hironaka's} theorem,'' \emph{{IEEE} Trans. Inf. Theory}, vol.~59, no.~7, pp.
  4213--4229, Jul. 2013.

\bibitem{koliander13-10a}
\BIBentryALTinterwordspacing
G.~Koliander, E.~Riegler, G.~Durisi, and F.~Hlawatsch, ``Degrees of freedom of
  generic block-fading {MIMO} channels without a priori channel state
  information,'' \emph{{IEEE} Trans. Inf. Theory}, Oct. 2013, submitted for
  publication. [Online]. Available: \url{http://arxiv.org/abs/1310.2490}
\BIBentrySTDinterwordspacing

\bibitem{jindal10-10a}
N.~Jindal and A.~Lozano, ``A unified treatment of optimum pilot overhead in
  multipath fading channels,'' \emph{{IEEE} Trans. Commun.}, vol.~58, no.~10,
  pp. 2939-- 2948, Oct. 2010.

\bibitem{Meyr1998}
H.~Meyr, M.~Moeneclaey, and S.~Fechtel, \emph{Digital Communication Receivers:
  Synchronization, Channel Estimation and Signal Processing, 1st ed.}\hskip 1em
  plus 0.5em minus 0.4em\relax New York, NY, USA: John Wiley \& Sons, 1998.

\bibitem{gallager68a}
R.~G. Gallager, \emph{Information Theory and Reliable Communication}.\hskip 1em
  plus 0.5em minus 0.4em\relax New York, NY, U.S.A.: Wiley, 1968.

\bibitem{gelfand57-a}
I.~M. Gel'fand and A.~M. Yaglom, ``Calculation of the amount of information
  about a random function contained in another such function,'' \emph{Uspekhi
  Mat. Nauk}, vol.~12, no.~1, pp. 3--52, 1957, {English transl.}, Amer. Math.
  Soc. Transl., Ser. 2, Vol. 12, 1959, pp. 199-246.

\bibitem{wyner66-03a}
A.~D. Wyner, ``The capacity of the band-limited {Gaussian} channel,''
  \emph{Bell Syst. Tech.~J.}, vol.~45, no.~3, pp. 359--395, Mar. 1966.

\bibitem{ghozlan13-07a}
H.~Ghozlan and G.~Kramer, ``On {Wiener} phase noise channels at high
  signal-to-noise ratio,'' in \emph{Proc. IEEE Int. Symp. Inf. Theory (ISIT)},
  Istanbul, Turkey, Jul. 2013, pp. 2279--2283.

\bibitem{koch10-11a}
T.~Koch and A.~Lapidoth, ``Increased capacity per unit-cost by oversampling,''
  in \emph{Proc. IEEE 26th Convention of Electrical and Electronics Engineers
  in Israel (IEEEI)}, Eilat, Israel, Nov. 2010, pp. 684 --688.

\bibitem{Gilbert93}
E.~N. Gilbert, ``Increased information rate by oversampling,'' \emph{{IEEE
  Trans. Inf. Theory}}, vol.~39, no.~6, pp. 1973--1976, 1993.

\bibitem{Shamai94}
S.~Shamai, ``Information rates by oversampling the sign of a bandlimited
  process,'' \emph{{IEEE Trans. Inf. Theory}}, vol.~40, no.~4, pp. 1230--1236,
  1994.

\bibitem{durisi11-08a}
G.~Durisi and H.~B\"{o}lcskei, ``High-{SNR} capacity of wireless communication
  channels in the noncoherent setting: A primer,'' \emph{Int. J. Electron.
  Commun. (AE{\"U})}, vol.~65, no.~8, pp. 707--712, Aug. 2011, invited paper.

\bibitem{NeeserMassey1993}
F.~D. Neeser and J.~L. Massey, ``Proper complex random processes with
  applications to information theory,'' \emph{{IEEE Trans. Inf. Theory}},
  vol.~39, no.~4, pp. 1293--1302, Jul. 1993.

\bibitem{hojo85}
R.~A. Horn and C.~R. Johnson, \emph{{M}atrix {A}nalysis}.\hskip 1em plus 0.5em
  minus 0.4em\relax Cambridge, UK: Cambridge Univ. Press, 1985.

\bibitem{Cover91}
T.~M. Cover and J.~A. Thomas, \emph{{E}lements of {I}nformation {T}heory},
  2nd~ed.\hskip 1em plus 0.5em minus 0.4em\relax New York, NY: Wiley, 2006.

\bibitem{alcami12}
B.~Alexeev, J.~Cahill, and D.~G. Mixon, ``Full spark frames,'' \emph{J Fourier
  Anal Appl}, vol.~18, no.~6, pp. 1167--1194, Dec. 2012.

\end{thebibliography}

%%%%%%%%%%%%%%%%%%%%%%%%%%%%%

%\pagebreak

\end{document}